\documentclass{appolb}

\usepackage{graphicx}
\usepackage{epsfig}
\usepackage{rotating}
\usepackage{amsmath}
\usepackage{bm,amsfonts,amssymb}
\usepackage{comment}

\def\be{\begin{equation}}
\def\ee{\end{equation}}
\def\bea{\begin{eqnarray}}
\def\eea{\end{eqnarray}}
\def\bdm{\begin{displaymath}}
\def\edm{\end{displaymath}}

\newcommand{\barr}{\begin{array}}
\newcommand{\earr}{\end{array}}

\newcommand{\dsc}{$D_{s1}(2536)$}
\newcommand{\dsd}{$D_{s1}(2460)$}
\newcommand{\dc}{$D_{1}(2420)$}
\newcommand{\dd}{$D_{1}(2430)$}
\newcommand{\ssz}{${}^{1\!}S_0$}
\newcommand{\tso}{${}^{3\!}S_1$}
\newcommand{\tdo}{${}^{3\!}D_1$}
\newcommand{\tpz}{${}^{3\!}P_0$}
\newcommand{\tpo}{${}^{3\!}P_1$}
\newcommand{\tpt}{${}^{3\!}P_2$}
\newcommand{\sdt}{${}^{1\!}D_2$}
\newcommand{\tdt}{${}^{3\!}D_2$}
\newcommand{\tft}{${}^{3\!}F_2$}
\newcommand{\otpo}{$1\,{}^{3\!}P_1$}
\newcommand{\ttpo}{$2\,{}^{3\!}P_1$}
\newcommand{\spo}{${}^{1\!}P_1$}
\newcommand{\ospo}{$1\,{}^{1\!}P_1$}
\newcommand{\tspo}{$2\,{}^{1\!}P_1$}
\newcommand{\otpt}{$1\,{}^{3\!}P_2$}
\newcommand{\ttpt}{$2\,{}^{3\!}P_2$}
\newcommand{\otpz}{$1\,{}^{3\!}P_0$}
\newcommand{\ttpz}{$2\,{}^{3\!}P_0$}

\newcommand{\tssz}{$2\,{}^{1\!}S_0$}
\newcommand{\hssz}{$3\,{}^{1\!}S_0$}
\newcommand{\ttso}{$2\,{}^{3\!}S_1$}
\newcommand{\htso}{$3\,{}^{3\!}S_1$}
\newcommand{\otdo}{$1\,{}^{3\!}D_1$}
\newcommand{\ttdo}{$2\,{}^{3\!}D_1$}
\newcommand{\osdt}{$1\,{}^{1\!}D_2$}
\newcommand{\tsdt}{$2\,{}^{1\!}D_2$}
\newcommand{\otdt}{$1\,{}^{3\!}D_2$}
\newcommand{\ttdt}{$2\,{}^{3\!}D_2$}
\newcommand{\jpsi}{J\!/\!\psi}

\newcommand{\bsm}[1]{\boldsymbol{#1}}
\begin{document}
\title{Meson spectroscopy: \\ too much excitement and too few
excitations\thanks{Presented by G.~Rupp at the Workshop ``Excited QCD 2012'',
Peniche, Portugal, 7--11 May 2012.}%
}
\author{George Rupp$^*$, Susana Coito,
\address{
Centro de F\'isica das Interac\c c\~oes Fundamentais, Instituto Superior
T\'ecnico, Technical University of Lisbon, P-1049-001 Lisboa, Portugal}
\and
Eef van Beveren
\address{
Centro de F\'{\i}sica Computacional, Departamento de F\'{\i}sica,
Universidade de Coimbra, P-3004-516 Coimbra, Portugal}
}

\maketitle
\renewcommand*{\thefootnote}{\fnsymbol{footnote}}
\setcounter{footnote}{2}

\begin{abstract}
We briefly review the general status of meson spectroscopy, especially in
light of the often made claim that there are too many observed resonances
to be accounted for as $q\bar{q}$ states. Also, the adequacy of the usual 
Coulomb-plus-linear, alias ``funnel'', confining potential for reproducing
the experimental spectra of light, heavy-light, and heavy mesons is critically
analysed. Thus, many serious discrepancies are observed and discussed. As
possible causes, we suggest the neglect of unitarisation and other
coupled-channel effects, as well as the deficiency of the funnel potential
itself. In order to illustrate our alternative, ``unquenched'' approach, we
present some recent examples of successfully described puzzling mesonic
enhancements and resonances, such as the charmonium states $X(4260)$ and
$X(3872)$, as well as the axial-vector charmed mesons \dc, \dd, \dsc, and
\dsd.
\end{abstract}

\PACS{14.40.-n, 13.25.-k, 12.40.Yx, 11.80.Gw}
\section{Brief review of the meson spectrum}
Contrary to widespread belief, there are not too many observed
\cite{PRD86p010001} mesonic resonances to be accounted for by normal
quark-antiquark states, which is the usually invoked argument to justify the
introduction of exotic, i.e., non-$q\bar{q}$, configurations. The reasons why
newly detected mesonic enhancements often seem to be incompatible
with $q\bar{q}$ states can be manifold:
\begin{enumerate}
\item
the underlying confining potential may be different from what is generally
taken for granted, as often demonstrated by us
(see e.g.\ Ref.~\cite{POSHQL2010p3});
\item
mass shifts due to unitarisation (or ``unquenching'') are mostly neglected
or underestimated (see e.g.\ Ref.~\cite{PRL91p012003});
\item
unitarisation sometimes even yields extra, dynamically generated resonances,
which nevertheless have a $q\bar{q}$ source (see e.g.\ Ref.~\cite{ZPC30p615});
\item
the opening of strong decay thresholds generally distorts the line shapes of
nearby resonances, or can even by itself give rise to enhancements that
look like resonances (see e.g.\ Ref.~\cite{PRD80p074001});
\item
large inelasticity effects between observed OZI-forbidden decays and
non-observed OZI-allowed ones can lead to signal depletion at true resonances
and/or thresholds, resulting in non-resonant apparent enhancements in between
(see e.g.\ Ref.~\cite{PRL105p102001}).
\end{enumerate}
In this talk, an short assessment will be made of the status of meson
spectroscopy. As theoretical benchmark we shall take what many consider a
kind of ``standard model'' of mesons, viz.\ the celebrated and topcited
relativised model of Godfrey \& Isgur (GI) \cite{PRD32p189}, which features 
the ususal Coulomb-plus-linear confinement forces, also called ``funnel''
potential, and an explicit one-gluon-exchange term generating spin-spin and
spin-orbit splittings. After showing numerous discrepancies between the
predictions of the GI model and data \cite{PRD86p010001}, we shall review an
alternative description of three controversial mesonic systems, namely the
$X(4260)$ \cite{PRD86p010001} and $X(3872)$ \cite{PRD86p010001} charmonium
structures, as well as the \dsd\ \cite{PRD86p010001} charmed-strange meson.
These states cannot be described correctly by the GI or any other quenched
$q\bar{q}$ model, giving rise to the usual ``poor-man's'' explanation in terms
of exotics or crypro-exotics.  Below, we shall show how our unitarised
Resonance-Spectrum Expansion \cite{AOP324p1620} manages to explain
$X(4260)$ \cite{PRL105p102001} as a non-resonant enhancement, and reproduce
the true resonances $X(3872)$ \cite{EPJC71p1762} 
and \dsd\ \cite{PRD84p094020}, together \cite{PRD84p094020} with the other
axial-vector charmed mesons \dsc, \dd, and \dc\ \cite{PRD86p010001}.

\subsection{Observed meson spectum and Godfrey-Isgur \cite{PRD32p189} model}
The GI \cite{PRD32p189} quark model for mesons is still referred to very
frequently for comparison when new resonances are observed or other models
make predictions. This is understandable in view of the GI model's
completeness in predicting meson spectra for almost any desired flavour
combinations and quantum numbers, besides the employment of the widely
accepted funnel potential. Thus, it appears opportune to briefly reassess this
27-year-old model in the light of present-day meson spectra as interpreted
by the PDG \cite{PRD86p010001} collaboration, identifying some of the 
outstanding problems.
\subsubsection{Light-quark isoscalar mesons}
\begin{itemize}
\item
$0^{++}$/\tpz: \\
Lowest GI scalar $\sim\!600$ MeV heavier than $\bsm{f_0(600)}$\footnote
{Henceforth, we shall print the states included in the PDG Summary Table
\cite{PRD86p010001} in boldface.} (alias $\bsm{\sigma}$); \\
GI $s\bar{s}$ scalar almost 400 MeV heavier than
$\bsm{f_0(980)}$.
\item
$2^{++}$/\tpt-\tft: \\
PDG listings report 6 likely $n\bar{n}$ ($n=u,d$) states up to
$\approx\!2.15$~GeV, viz.\ $\bsm{f_2(1270)}$, $f_2(1565)$,
$f_2(1640)$, $f_2(1810)$, $f_2(1910)$, and $f_2(2150)$, whereas GI only
predict 3. In probably dominant $s\bar{s}$ sector, PDG also lists 6 states
up to $\approx\!2.35$ GeV: $f_2(1430)$, $\bsm{f_2^\prime(1525)}$,
$\bsm{f_2(1950)}$, $\bsm{f_2(2010)}$, $\bsm{f_2(2300)}$, and
$\bsm{f_2(2340)}$, while GI again only predict 3. \\[1mm]
Note: some PDG $f_2$ states may not be resonances \cite{PR397p257},
but $f_2(1565)$ looks reliable. Also,
PDG: $m(\mbox{\ttpt})-m(\mbox{\otpt})\approx300$~MeV; 
GI: $m(\mbox{\ttpt})-m(\mbox{\otpt})=540$~MeV. \\
For unknown reasons, PDG omits $f_2(1565)$ from the Summary Table.
\item
$1^{+-}$/\spo: \\
PDG $n\bar{n}$ entries: $\bsm{h_1(1170)}$,
$h_1(1595)$; \\
GI predict: $h_1(1220)$ (\ospo), $h_1(1780)$ (\tspo).
\end{itemize}
\subsubsection{Light-quark isovector mesons}
\begin{itemize}
\item
$0^{++}$/\tpz: \\
PDG entries: $\bsm{a_0(980)}$, $\bsm{a_0(1450)}$; \\
GI: $a_0(1090)$ (\otpz), $a_0(1780)$ (\ttpz).
\item
$1^{++}$/\tpo: \\
PDG entries: $\bsm{a_1(1260)}$, $a_1(1640)$; \\
GI: $a_1(1240)$ (\otpo), $a_1(1820)$ (\ttpo).
\item
$2^{++}$/\tpt: \\
PDG entries: $\bsm{a_2(1320)}$, $a_2(1700)$; \\
GI: $a_2(1310)$ (\otpt), $a_2(1820)$ (\ttpt).
\item
$1^{--}$/\tso-\tdo: \\
PDG entries: $\bsm{\rho(1450)}$, $\rho(1570)$,
$\bsm{\rho(1700)}$, $\rho(1900)$;  \\
GI: $\rho(1450)$ (\ttso), $\rho(1660)$ (\otdo),
$\rho(2000)$ (\htso), $\rho(2150)$ (\ttdo). \\
Note: a recent analytic $S$-matrix analysis \cite{NPA807p145} 
arrived at assignments quite different from
both PDG and GI: $\rho(1250)$, $\rho(1470)$, $\rho(1600)$,
$\rho(1900)$. Also, it concluded that only $\rho(1250)$
and $\rho(1600)$ are crucial to describe the phase shifts,
whereas $\rho(1900)$ and, to a lesser extent, $\rho(1470)$
improve the inelasticity. For mysterious reasons, PDG conceals
$\rho(1250)$ under the $\bsm{\rho(1450)}$ entry \cite{PRD86p010001}.
\end{itemize}
\subsubsection{Strange mesons}
\begin{itemize}
\item
$0^{-}$/\ssz: \\
PDG entries: $K(1460)$, $K(1830)$; \\
GI: $K(1450)$ (\tssz), $K(2020)$ (\hssz).
\item
$0^{+}$/\tpz: \\
PDG entries: $K_0^\ast(800)$, $\bsm{K_0^\ast(1430)}$, $K_0^\ast(1950)$; \\
GI: $K_0^\ast(1240)$ (\otpo), $K_0^\ast(1890)$ (\ttpo). 
\item
$1^{-}$/\tso-\tdo:
PDG entries: $\bsm{K^\ast(1410)}$, $\bsm{K^\ast(1680)}$; \\
GI: $K^\ast(1580)$ (\ttso), $K^\ast(1780)$ (\otdo).
\item
$1^{+}$/\tpo-\spo: \\
PDG entries: $\bsm{K_1(1270)}$, $\bsm{K_1(1400)}$,
$K_1(1650)$; \\
GI: $K_1(1340)$ (\ospo), $K_1(1380)$ (\otpo), $K_1(1900)$ (\tspo),
$K_1(1930)$ (\ttpo).
\item
$2^{-}$/\sdt-\tdt: \\
PDG entries: $K_2(1580)$, $\bsm{K_2(1770)}$,
$\bsm{K_2(1820)}$, $K_2(2250)$; \\
GI: $K_2(1780)$ (\osdt), $K_2(1810)$ (\otdt),
$K_2(2230)$ (\tsdt), $K_2(2260)$ (\ttdt).
\end{itemize}
\subsubsection{Summary of light mesons}
The light-meson spectrum \cite{PRD86p010001} appears to favour radial
splittings that are considerably smaller than those predicted by the GI
and similar funnel models, as well as lattice QCD \cite{PRD83p111502}.
There is no indication that some of the observed resonances might be
crypto-exotics, so that no excess of states can be claimed. On the other
hand, missing states in e.g.\ the strange and $\phi$ sectors make definite
conclusions on the confining force even more difficult.
\subsubsection{Charmed mesons}
Especially the scalar $\bsm{D_{s0}^\ast(2317)}$ but also the axial-vector
$\bsm{D_{s1}(2460)}$ come out too heavy in the GI model (also see below).
For the rest, too many quark-model states have not been observed so far, 
which hardly allows any feedback concerning the confining potential.
\subsubsection{Charmonium}
In recent years, the PDG listings have been invaded by a plague of
charmonium-like, so-called ``$X$'' states, several of which may not be
resonances at all, but rather threshold \cite{PRD80p074001} or depletion
\cite{PRL105p102001} effects (see e.g.\ $\bsm{X(4260)}$
\cite{PRL105p102001} below).
Moreover, genuine charmonium states like $\bsm{X(3872)}$ may be shifted
considerably \cite{EPJC71p1762}, turning a correct assignment into a much
more difficult task than simply checking the mass predictions of one's
favou\-rite quenched quark model. As for the ``regular'' charmonium spectrum,
too many spectroscopic states have evaded observation so far to allow a better
understanding of confinement. On the other hand, clear indications of highly
excited radial vector states \cite{POSHQL2010p3} have been systematically
ignored by other model builders and the experimental groups themselves.
\subsubsection{Bottom mesons}
Here, the scarcity of observed \cite{PRD86p010001} excited states strongly
hampers any significant contribution to meson spectroscopy.
\subsubsection{Bottomonium}
As for bottomonium, a correct spectroscopic assignment of
$\bsm{\Upsilon(10580)}$, $\bsm{\Upsilon(10860)}$, and $\bsm{\Upsilon(11020)}$
\cite{PRD86p010001} is crucial to understand the interplay of confinement and
coupled channels above the open-bottom threshold. The usual interpretation of
these states as $\Upsilon(4S)$, $\Upsilon(5S)$, and $\Upsilon(6S)$,
respectively, suffers from serious problems (see Ref.~\cite{POSHQL2010p3}
and references therein). Also, future experiments at LHC must certainly improve
\cite{ARXIV12041984} on the resolution achieved in Ref.~\cite{PRL108p152001}
for any real advancement in spectroscopy.
\section{Non-resonant charmonium enhancement $X(4260)$}
\begin{figure}[t]
\begin{tabular}{c}
\hspace*{0.2cm}
\includegraphics[scale=0.60,trim=0cm 0.4cm 0cm 0.4cm, clip=true]
{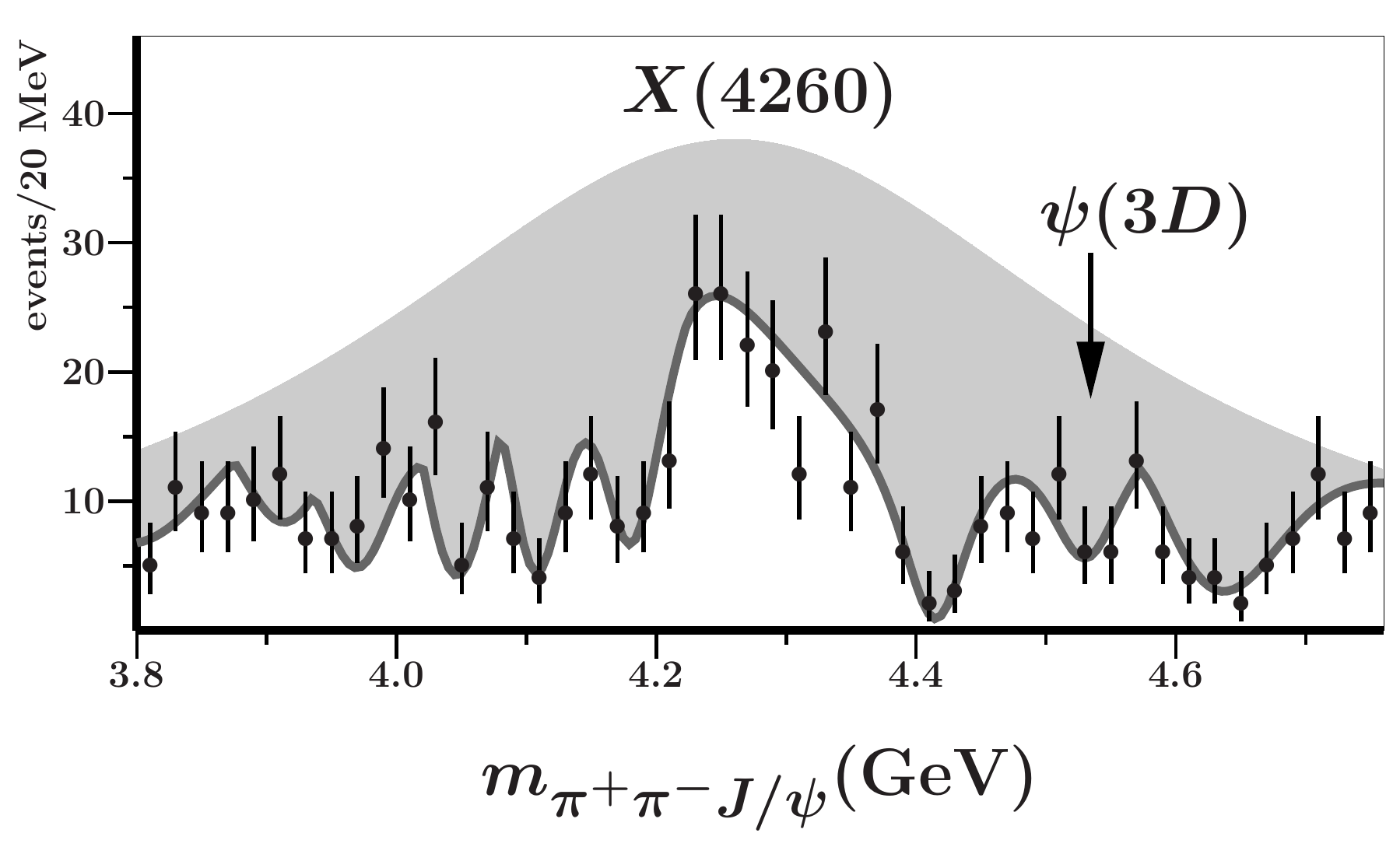}
\end{tabular}
\caption{Non-resonant $\bsm{X(4260)}$ modelled as a very broad structure
(shaded area) depleted by $c\bar{c}$ resonances and open-charm thresholds.
Effect of new charmonium state $\psi(3D)$ is clearly visible.
Data are from Ref.~\cite{PRL95p142001}; also see
Refs.~\cite{PRL105p102001,PPNP67p449}.}
\label{octopsi}
\end{figure}

The $\bsm{X(4260)}$ vector charmonium enhancement \cite{PRD86p010001},
discovered \cite{PRL95p142001} in $\pi^+\pi^-\jpsi$ data, is puzzling
because of its awkward mass and the non-obser\-vation of open-charm decay
channels. This has led to various exotic or molecular model explanations (see
Ref.~\cite{PRL105p102001} for some references). However, the $\bsm{X(4260)}$
data also display a conspicuous dip precisely at the mass of the established
$\bsm{\psi(4415)}$, as well as the absence of peaks corresponding to other
known $c\bar{c}$ states. These usually ignored yet very peculiar features
can be understood by assuming \cite{PRL105p102001} a very broad,
$\bsm{\sigma}$-like, $\pi^+\pi^-$ distribution centred around 4.26 GeV,
but depleted at the energies of $c\bar{c}$ resonances, including a new
$\psi(3D)$ state at about 4.53~GeV, as well as open-charm threshold openings
(see Fig.~\ref{octopsi}). Thus, dominant, OZI-allowed processes reveal
themselves as a kind of ``mirror images'' in the --- OZI-suppressed ---
$\pi^+\pi^-\jpsi$ data. For further details, see Ref.~\cite{PRL105p102001}.

\section{$\bsm{X(3872)}$ as a unitarised $1^{++}$ $c\bar{c}$ state}
\begin{figure}[t]
\begin{tabular}{cc}
\hspace*{-12mm}
\includegraphics[scale=0.40,trim=0cm -1.6cm 0cm 0.6cm,clip=true]
{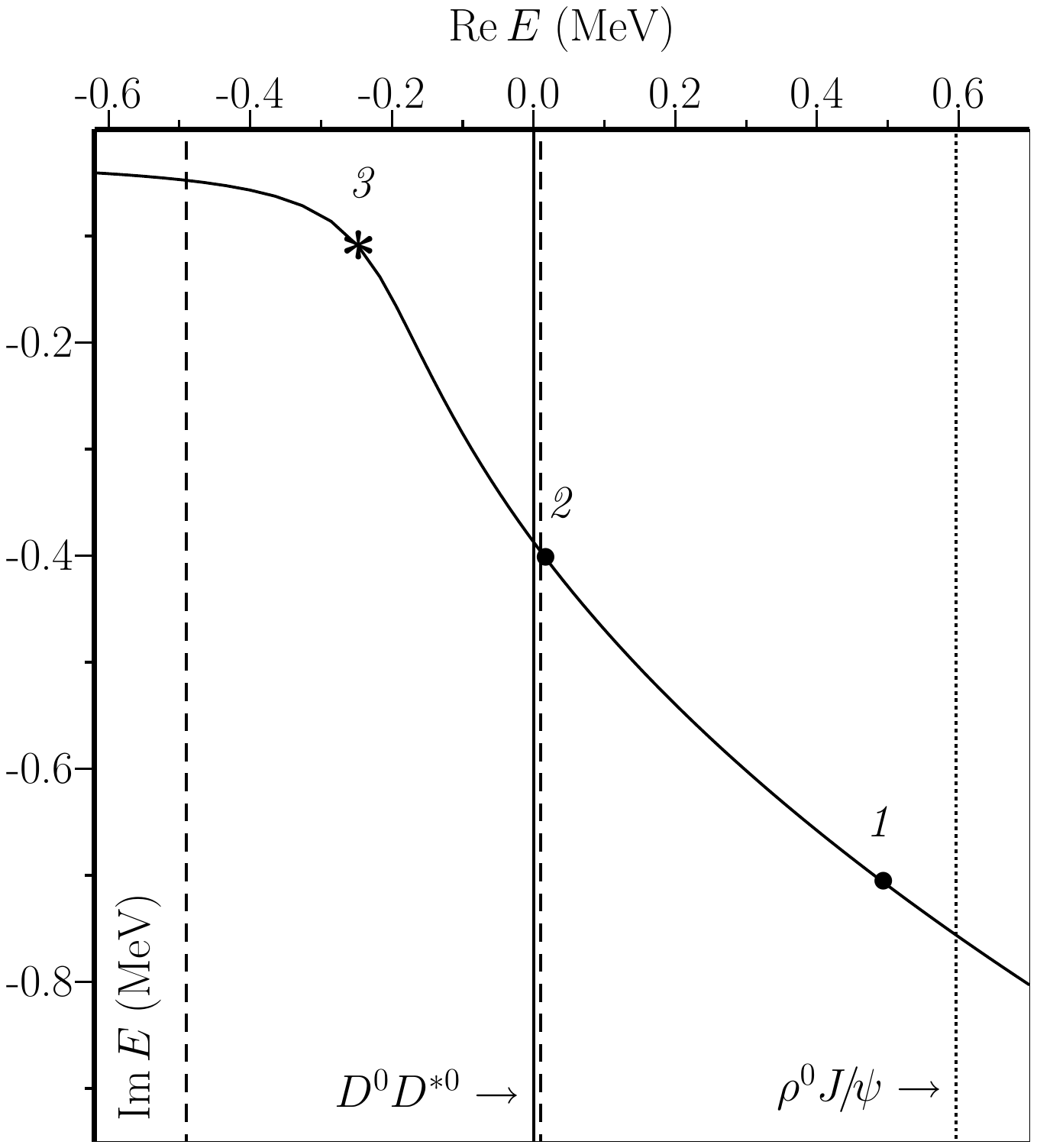} & \hspace*{-2.7cm}
\includegraphics[scale=0.40]{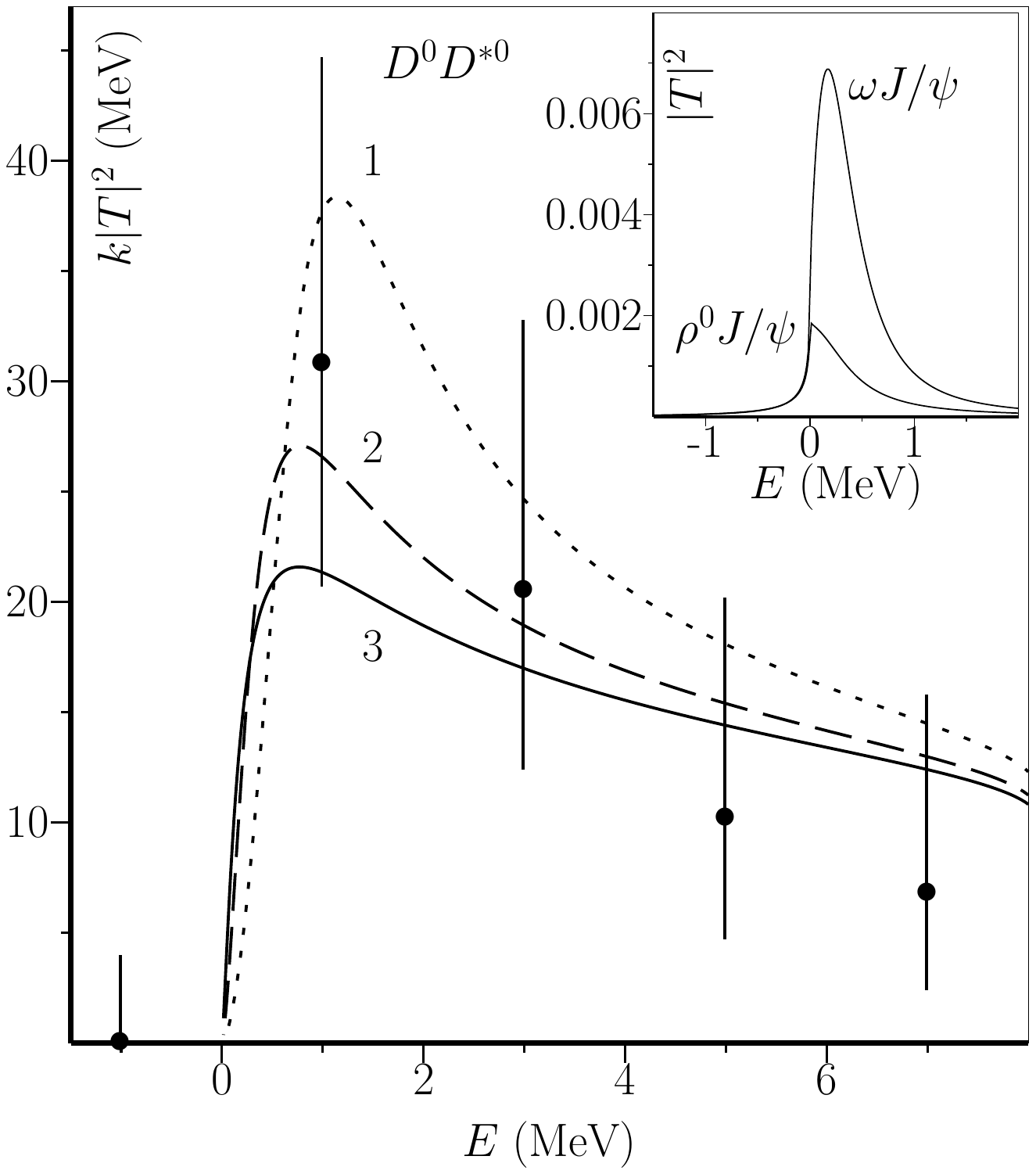}
\mbox{ } \\[-5.0cm]
\end{tabular}
\caption{Left: $\bsm{X(3872)}$ pole trajectory as a function of overall
coupling $\lambda$; dashed lines delimit allowed \cite{PRD86p010001} pole
range. Right: $D^0D^{\ast0}$ amplitude with data \cite{PRL91p262001};
inset: relative comparison of $\rho^0\jpsi$ and $\omega\jpsi$ amplitudes.
Also see Refs.~\cite{EPJC71p1762,PPNP67p449}. \vspace*{-3mm}}
\label{x3872}
\end{figure}

The $\bsm{X(3872)}$ charmonium-like state was discovered \cite{PRL91p262001}
as a $\pi^+\pi^-\jpsi$ enhancement in the decay
$B^\pm\to K^\pm\pi^+\pi^-\jpsi$, and later also observed in the hadronic
channels $\rho^0\jpsi$, $\omega\jpsi$, $D^0\bar{D}^0\pi^0$, and
$D^0\bar{D}^{\ast0}$ \cite{PRD86p010001}. The low mass of $\bsm{X(3872)}$
and its remarkable proximity to the $D^0D^{\ast0}$ threshold has given rise
to many exotic or molecular interpretations (see Ref.~\cite{EPJC71p1762}
for some references). However, we have shown \cite{EPJC71p1762} that
$\bsm{X(3872)}$ is perfectly compatible with a unitarised \ttpo\ $c\bar{c}$
state, but strongly mass-shifted and with a large $D^0D^{\ast0}$ component in
the wave function \cite{CRB2012}. Figure~\ref{x3872} displays the
$\bsm{X(3872)}$ pole trajectory near the $D^0D^{\ast0}$ threshold, as well
as a comparison of predicted amplitudes with data. For more information,
see Refs.~\cite{EPJC71p1762,CRB2012}.
\section{Understanding the axial-vector charmed mesons}
\begin{figure}[b]
\begin{tabular}{cc}
\hspace*{-11mm}
\includegraphics[scale=0.40,trim=0cm 11.5cm 0cm 0.65cm,clip=true]
{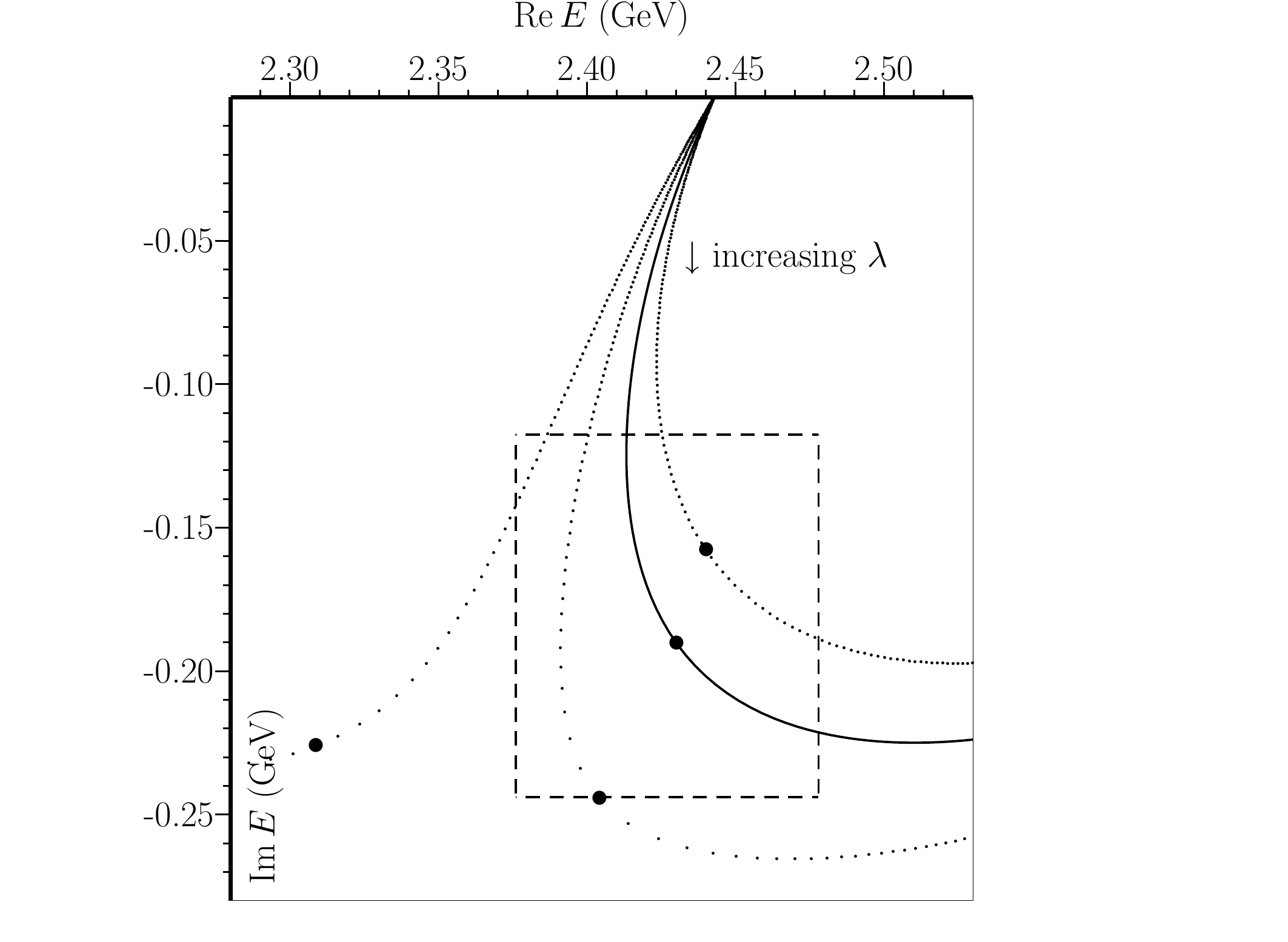} &
\hspace*{-2.7cm}
\includegraphics[scale=0.40,trim=0cm 11.5cm 0cm 0.65cm,clip=true]
{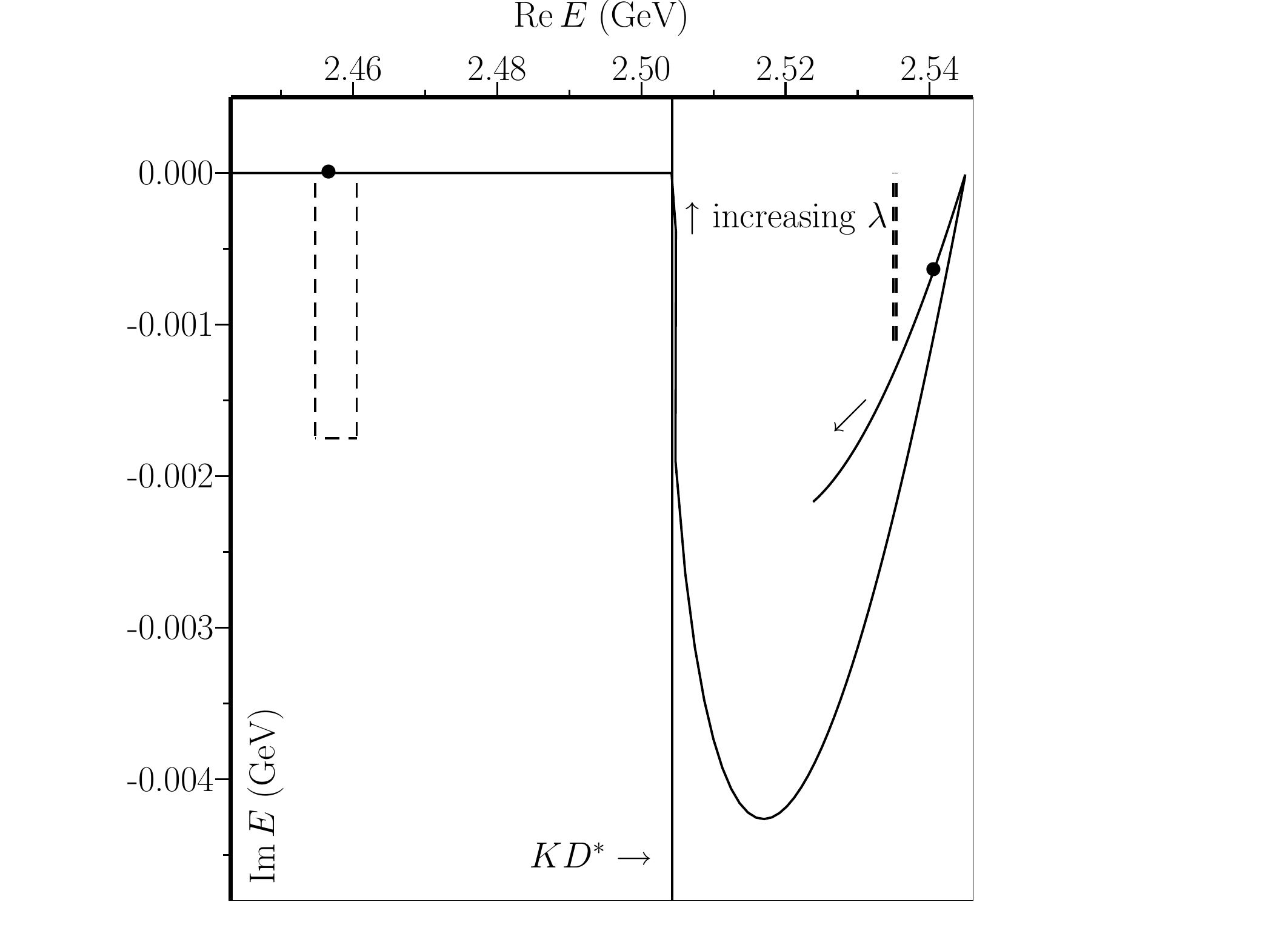} 
\end{tabular}
\caption{Left: pole trajectories of \dd\ as a function of coupling $\lambda$,
for different radii $r_0$;
right: trajectories of $\bsm{D_{s1}(2460)}$ and $\bsm{D_{s1}(2536)}$, as a
function of $\lambda$, for $r_0\!=\!3.12$~GeV$^{-1}$.
Left and right: dashed box, rectangle, and strip delimit experimental
\cite{PRD86p010001} \dd, $\bsm{D_{s1}(2460)}$ resp.\ $\bsm{D_{s1}(2536)}$ 
pole; see Refs.~\cite{PRD84p094020,PPNP67p449}.}
\label{axialcharm}
\end{figure}

One of the major puzzles in open-charm spectroscopy is the approximate mass
degeneracy of the axial-vector (AV) charmed resonances $\bsm{D_1(2420)}$ and
\dd, with the former being relatively narrow (20--25~MeV) and the latter
very broad ($\sim\!400$~MeV). On the other hand, the AV charmed-strange mesons
$\bsm{D_{s1}(2536)}$ and $\bsm{D_{s1}(2460)}$ lie 76~MeV apart, whereas their
widths are both very small (0--3~MeV). We have recently explained
\cite{PRD84p094020} this odd pattern of masses and widths by coupling the
various bare AV charmed states to their dominant decay channels. Unquenching
then not only moves the poles where they belong in the complex energy plane
(see Fig.~\ref{axialcharm}), but even generates the dynamical mixing of
\tpo\ and \spo\ components required to describe the data. See further
Ref.~\cite{PRD84p094020}.

\section{Conclusions}
Meson spectroscopy has progressed only marginally over the past decade when
judged on the obtained feedback concerning the confinement and decay
mechanisms in QCD, despite the observation of many new and exciting resonances.
Blame is to be put on both theorists and experimentalists, on the former ones
because of their obsession with the funnel potential and exotics, and on the
latter for failing to produce data with much higher resolution and carry 
out more systematic studies, especially partial-wave analyses \cite{PR397p257},
in various sectors of the meson spectrum. The clear evidence of large effects
from unquenching, as e.g.\ the mass shifts of $\bsm{D_{s0}^\ast(2317)}$
\cite{PRL91p012003} and $\bsm{X(3872)}$ \cite{EPJC71p1762}, the dynamically
generated light scalar nonet \cite{ZPC30p615,PRD86p010001}, or more
generally, due to threshold openings \cite{PRD80p074001}, should finally
convince people that modern meson spectroscopy must go beyond the traditional
quenched approaches like the GI \cite{PRD32p189} model, no matter how
pioneering the latter work was 27 years ago.

\end{document}